\documentstyle[epsfig,paspconf]{article}
\def\be{\begin{equation}}
\def\ee{\end{equation}}
\def\bea{\begin{eqnarray}}
\def\eea{\end{eqnarray}}

\def\etal{{\em et al.}            }

\def\mum{\mu {\rm m}}
\def\kms{{\rm \,km\,s}^{-1}}

\def\pster{{\rm \,ster}^{-1}}

\def\dg{^{\circ}}

\def\Jy{{\rm \,Jy}}
\def\MJy{{\rm \,MJy}}
\def\spose#1{\hbox to 0pt{#1\hss}}
\def\simlt{\mathrel{\spose{\lower 3pt\hbox{$\mathchar"218$}}
     \raise 2.0pt\hbox{$\mathchar"13C$}}}
\def\simgt{\mathrel{\spose{\lower 3pt\hbox{$\mathchar"218$}}
     \raise 2.0pt\hbox{$\mathchar"13E$}}}
\def\({\left(}
\def\){\right)}
\def\[{\left[}
\def\]{\right]}
\def\<{\left\langle}
\def\>{\right\rangle}

\def\ApJ{{\em Astrophys.~J.~}}

\def\IAU130{in {\em Large Scale Structures of the Universe}, IAU Symposium 130~}
\def\MN{{\em Mon.Not.R.astr.Soc.~}}

\def\edcomment#1{\iffalse\marginpar{\raggedright\sl#1\/}\else\relax\fi}
\reversemarginpar

\begin{document}
\title{The Behind The Plane Survey}
\author{Will Saunders$^\ast$, Kenton d'Mellow$^\ast$, Brent Tully$^\dag$ }
\affil{$^\ast$ IfA, University of Edinburgh. $^\dag$ IfA, University of Hawaii.}
\author{Bahram Mobasher$^\ast$, Steve Maddox$^\dag$}
\affil{$^\ast$ Blackett Lab, ICMST, University of London. $^\dag$  IoA, Cambridge University.}
\author{Will Sutherland$^\ast$, Esperanza Carrasco$^\dag$, George Hau$^\ddag$}
\affil{$^\ast$ Dept. of  Physics, Oxford University. $^\dag$ INAOE, Puebla. $^\ddag$ Universidad Catholica, Santiago.}
\author{ Dave Clements$^\ast$, Staveley-Smith$^\dag$}
\affil{$^\ast$ Dept. of  Physics, University of Cardiff. $^\dag$ Radiophysics, CSIRO.}

\begin{abstract}
We have extended the PSCz to cover 93\% of the sky, the practical limit for any
catalogue based on the Point Source Catalog (Beichman \etal 1988, ES). The galaxy identifications are complete, and the core of the Great Attractor is quantitatively mapped for the first time. We discuss the likely effect of the increased sky coverage on the IRAS dipole.
\end{abstract}

\section{Introduction}
\vspace{-5pt}
The 84\% sky coverage of the PSCz survey is effectively limited by the need to get, for every galaxy, an optical identification from sky survey plates. The fractional incompleteness in sky coverage translates directly into uncertainty in predicting the gravity dipole on the Local Group. The IRAS PSC data itself is reliable to much lower latitudes, although genuine galaxies are outnumbered by Galactic sources with similar IRAS properties. Previous attempts to go further into the Plane have either been restricted to the Arecibo declination range, or have relied on optical identifications from Sky Survey Plates. Because the extinction may be several magnitudes or more, they have inevitably suffered from progressive and unquantifiable incompleteness as a function of latitude. 

In 1994 we embarked on a program, parallel with the PSCz survey, to systematically identify low latitude IRAS galaxies wherever the PSC data allowed, using new near-infrared observations where necessary.

\section{Sky coverage and selection criteria}
\vspace{-5pt}
The mask consists of (a) the IRAS coverage gaps (3\% of the sky), (b) areas flagged as High Source Density at $60\mum$ (3\%), where the PSC processing was changed to ensure reliability at the expense of completeness, (c) areas flagged as High Source Density at $12$ or $25\mum$ on the basis that identifications would be impossible, and (d) areas with $I_{100} > 100 \MJy\pster$ (as in Rowan-Robinson \etal 1991), because of the impossible IRAS confusion and contamination by local sources. Our final sky coverage is 93\%.

The IRAS selection criteria were tightened from those used for the PSCz, in order to minimise the contamination by Galactic sources while still keeping most of the galaxies. The revised criteria were \\
\begin{tabular}{lcc} 
$S_{60}/S_{25}$ &$>$& $2$ \\
$S_{60}/S_{12}$ &$>$& $4$ \\
$S_{100}/S_{60}$ &$>$& $1$ \\
$S_{100}/S_{60}$ &$<$& $5$ \\
$CC_{60}$ &$=$& $A,B,C$
\end{tabular}

Upper limits were only used where they guaranteed inclusion or exclusion. 
At high latitudes, these criteria encompass the vast majority of galaxies, with incompleteness increasing from 3\% for nearby galaxies, to 6\% at $15,000\kms$.
At low latitudes, the incompleteness is higher because of corrupted fluxes. However, we know of just 18 galaxies at low latitude excluded by our selection criteria, compared with 1225 included. Unfortunately, the criteria efficiently exclude very nearby galaxies such as Dwingeloo I. From the source counts alone, it is clear that only one third of the sources are galaxies.

\section{Identifications}
\vspace{-5pt}
Many sources are immediately identifiable as galaxies from sky survey plates. Others are clearly Galactic. To identify the rest, we used Sky survey plates in all available bands, NVSS data for $\delta>-40\dg$, IRAS addscan profiles, Simbad and other literature data. For almost all sources still remaining unclassified, and also almost all sources with a faint ($B_j>19.5^m$) galaxy counterpart, we obtained $K'$ `snapshots' using the UHa $88'$$'$, UNAM 2.1m, ESO 2.2m, CTIO 1.5m and Las Campanas 1m telescopes, between 1994 and 1999.

In general, the $K'$ images allowed unambiguous identification as a galaxy or other (usually cirrus) source. Occasionally, there remained ambiguity between galaxies and buried YSO's, and more frequently a very faint and compact galaxy ID was overlooked, but  revealed by subsequent NVSS data. While hundreds of galaxies were found which are completely invisible on sky survey plates, in general some sort of optical counterpart is visible.

Our identification program gave us a total 1225 galaxies. They are plotted along with the PSCz survey in figure 1. Note the striking excess of galaxies in the Great Attractor region, $(l,b)=(325,-5)$, despite the low surface density of identified galaxies in the southern Galactic Plane. Note also that the northern sky is now essentially mapped right through the Plane, showing a bridge across linking Pisces and Perseus. We also see the Puppis and Ophiucus superclusters.

\begin{figure}
\centerline{\epsfig{figure=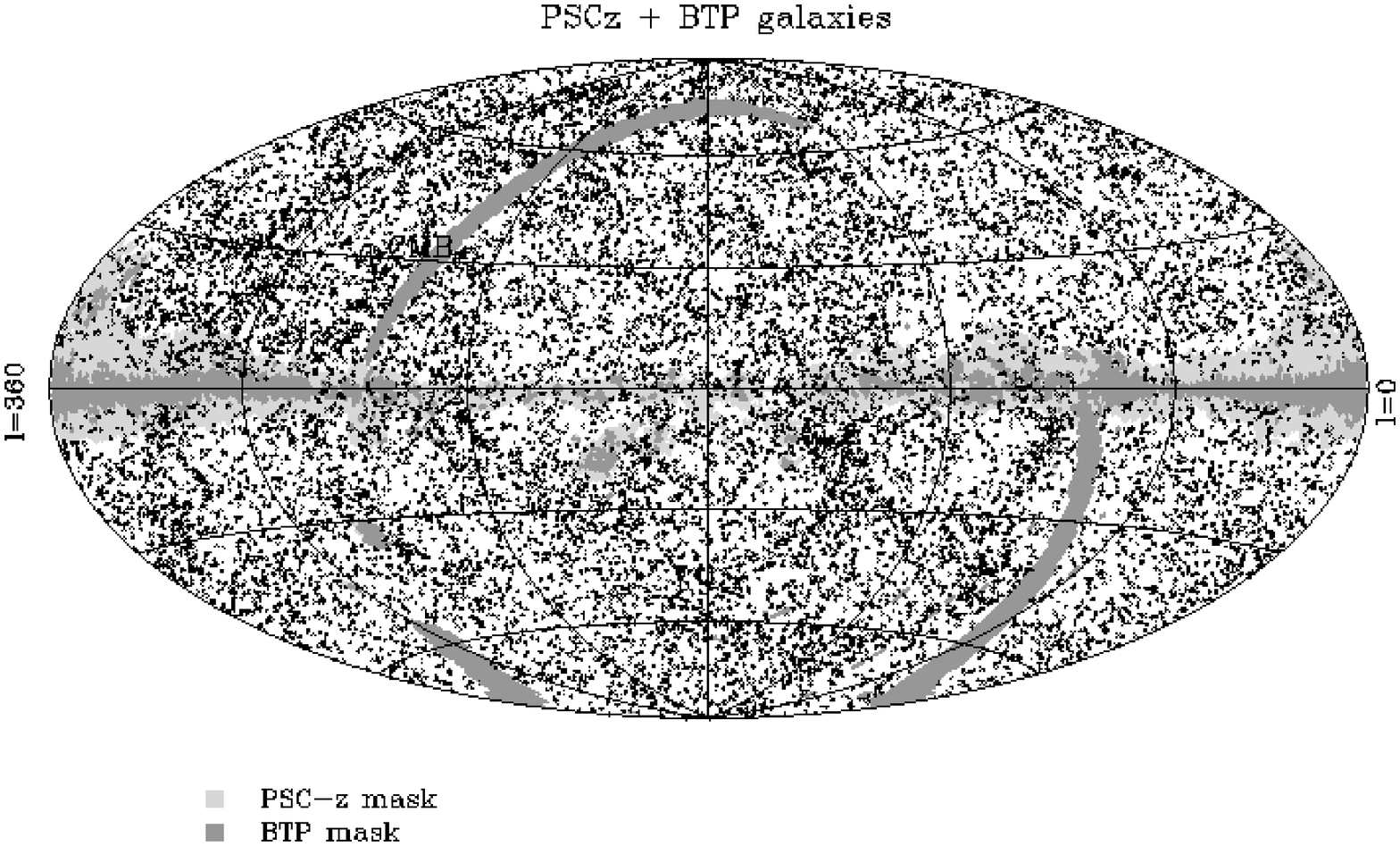,width=10cm,angle=-90}}
\vspace{-0.5in}
Figure 1. PSCz and BTP galaxy catalogues and masks in galactic coordinates.
\vspace{-5pt}
\end{figure}

\section{Completeness of the BTP survey}
\vspace{-5pt}
The nominal flux limit for the BTP survey is $0.6\Jy$, as for the PSCz. However, the source counts shown in figure 2 clearly show increasing incompleteness below $1Jy$ as a function of $I_{100}$. The primary causes of incompleteness in the PSC at low-latitudes are confusion and noise shadowing. Rather than trying to model these effects, we have simply defined a limit to which our source count completeness ($\sim 80\%$) is acceptable, as a function of $I_{100}$. We estimate the flux limit for reasonable completeness to be $S_{60lim}=0.5+I_{100}/(200\MJy\pster) \Jy$.
\begin{figure}
\centerline{\epsfig{figure=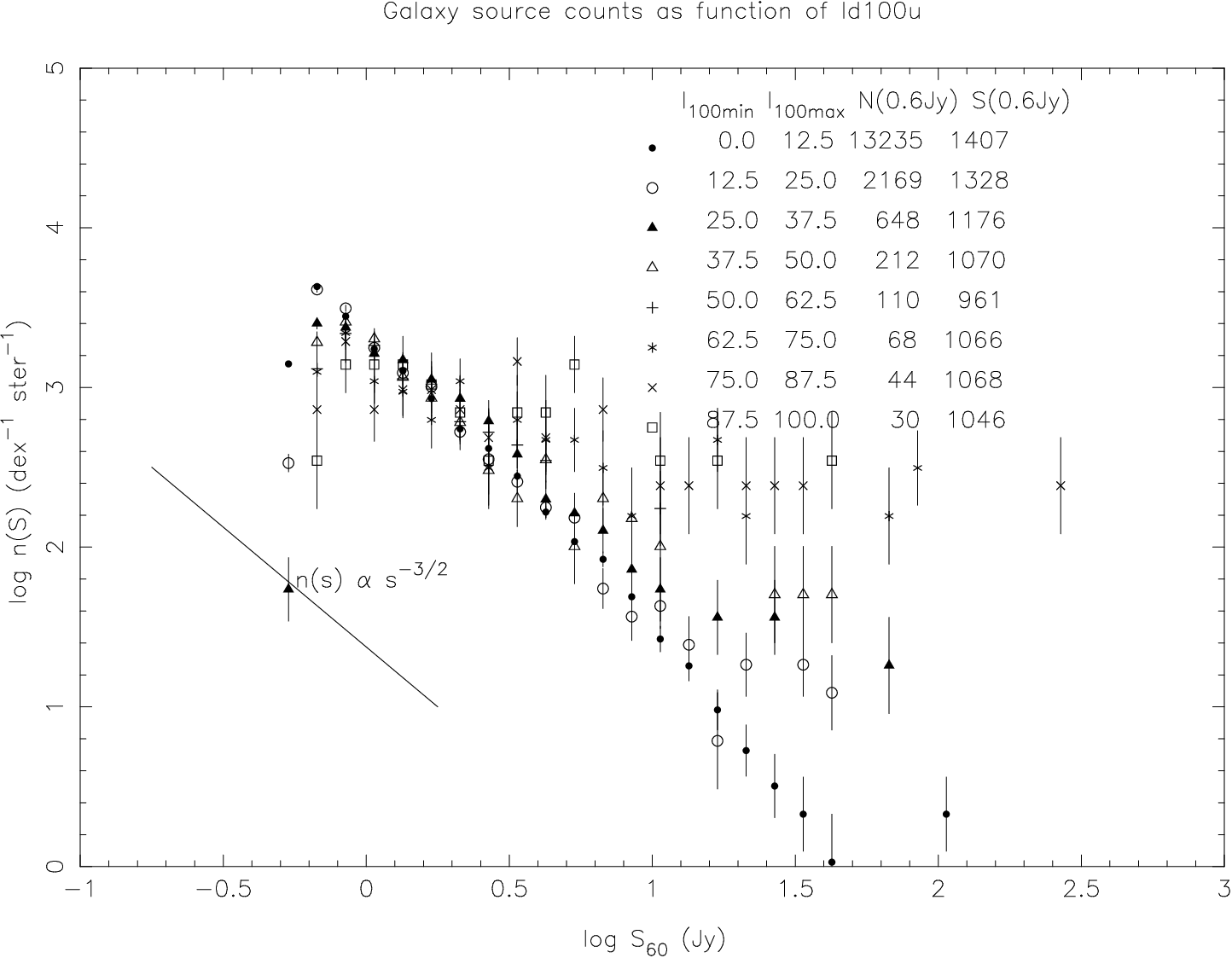,width=10cm,angle=-90}}
Figure 2. PSCz and BTP galaxy source counts as a function of $100\mum$ background.
\end{figure}

\section{2D surface density of galaxies}
\vspace{-5pt}
Combining the BTP survey with the PSCz allows to to map the surface distribution of IRAS galaxies across almost the whole sky. We have used the Fourier interpolation method described in Saunders \etal 1999, to interpolate across the residual mask. We have corrected for the effects of the incompleteness found above. The results are shown in figure 3; the power of the Great Attractor is now obvious, as easily the dominant feature in the map. There appears to be incompleteness larger than our above estimate within $10\dg$ of the Galactic Centre, and we are investigating possible causes.

\begin{figure}
\centerline{\epsfig{figure=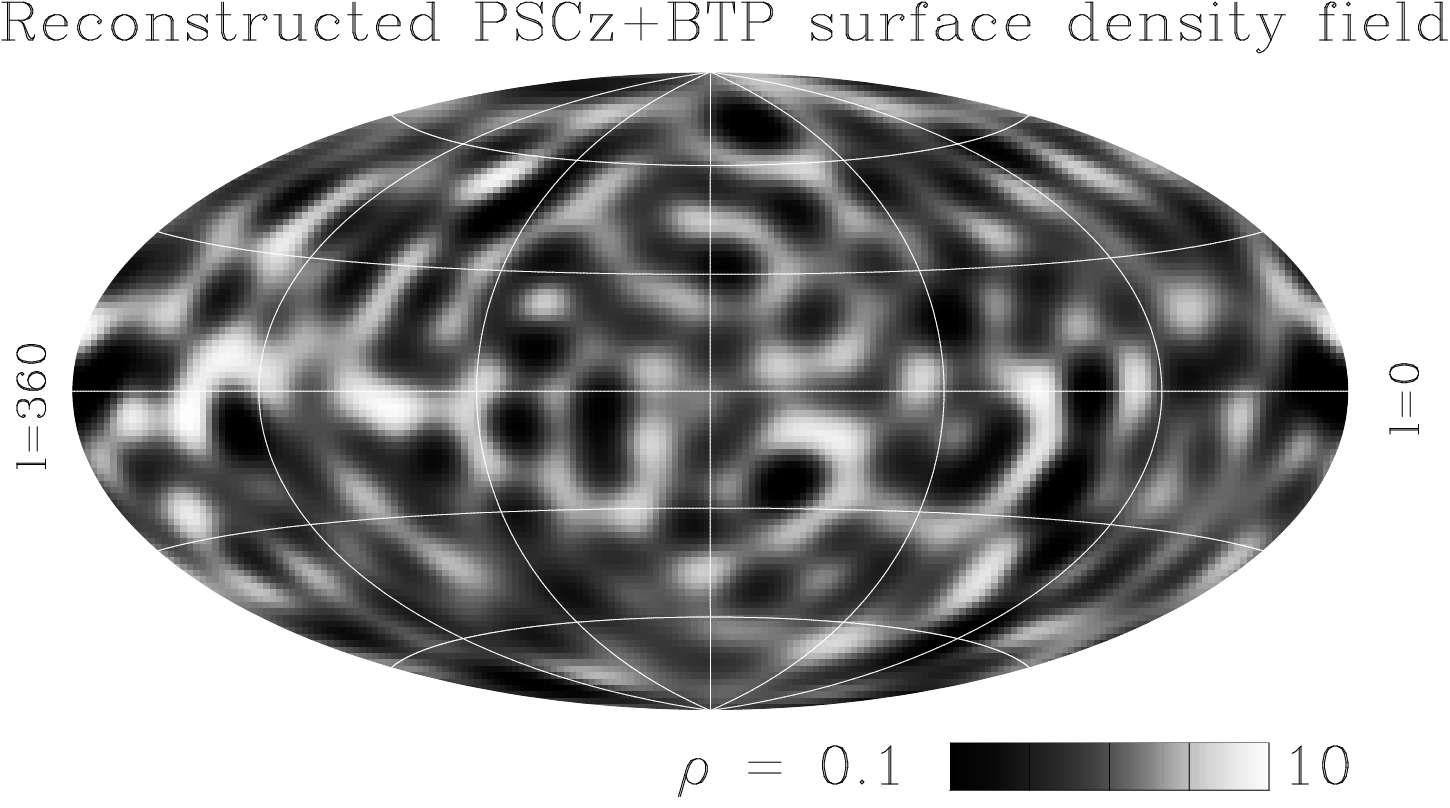,width=7.5cm,angle=-90}}
Figure 3. Surface density of PSCz+BTP galaxies in galactic coordinates.
\end{figure}

\section{Redshift aquisition}
\vspace{-5pt}
Complete spectroscopy for all sources is impracticable. However, in the PSCz, there is little contribution to the cumulative IRAS dipole beyond $15,000 \kms$, and this is an achievable completeness target; within this distance, and given the extinctions $A_R<4^m$, virtually all galaxies are accessible to optical or HI spectroscopy. About 40\% of the required redshifts were already known from previous surveys. The southern spectroscopy is now very nearly complete, using Parkes and the CTIO 1.5m and 4m telescopes. In the north, we have about 300 redshifts still to obtain.

\section{The effect on the IRAS dipole}
\vspace{-5pt}
The $20 \dg$ deviation between e.g. PSCz and CMB dipoles is much larger than the expected deviation for current, low $\Omega$ models; Strauss \etal (1992) found an rms misalignment of just $10\dg$.

The inclusion of the the BTP sample, and in particular the Great Attractor, will shift the IRAS dipole in both direction and magnitude. Given the relative positions of the GA, and the CMB and IRAS dipoles, the sense of this correction seems certain to improve the alignment of the IRAS and CMB dipoles, and to reduce the estimate of $\beta$.

\end{document}